\documentstyle[twocolumn,aps,epsfig]{revtex}
\begin{document}
\twocolumn[\hsize\textwidth\columnwidth\hsize\csname @twocolumnfalse\endcsname
\title{Local Charge Excesses in Metallic Alloys: a Local Field Coherent 
Potential Approximation Theory}

\author{Ezio Bruno\cite{mail}, 
Leon Zingales and Antonio Milici}

\address{Dipartimento di Fisica and Unit{\`{a}} INFM,
Universit{\`{a}} di Messina, Salita Sperone 31, 98166 Messina, Italy}
\date{\today}
\maketitle

\begin{abstract}
Electronic structure calculations performed on very large supercells have 
shown that the local charge excesses in metallic alloys are related through
simple linear relations to the local electrostatic field resulting from
distribution of charges in the whole crystal. 

By including local external fields in the isomorphous Coherent Potential 
Approximation theory, we develop a novel theoretical scheme in which the 
local charge excesses for random alloys can be obtained as the responses to 
local external fields. Our model maintains all the 
computational advantages of an isomorphous theory but allows for full charge 
relaxation at the impurity sites. Through applications to CuPd and CuZn 
alloys, we find that, as a general rule, non linear charge rearrangements 
occur at the impurity site as a consequence of the complex phenomena 
related with the 
electronic screening of the external potential. This nothwithstanding, 
we observe that linear relations hold between charge excesses and 
external potentials, in quantitative agreement with the mentioned supercell 
calculations, and well beyond the limits of linearity for any other site 
property.

\\
\\
\end{abstract}
]

\section{Introduction}
For many years the Coherent Potential Approximation (CPA)~\cite{Soven} has 
been widely used for calculating the electronic properties of random 
metallic alloys. There are many reasons for such a fortune: as it has been 
already pointed out, the CPA allows for a very careful determination of 
spectral properties~\cite{Abrikosov_cpa} and Fermi surfaces~\cite{physrep}. 
However, for the arguments discussed in the present paper, two other 
aspects of the CPA are particularly amenable: its simple and elegant 
formulation~\cite{Gyorffy} in terms of the multiple scattering 
theory~\cite{KKR} and the fact that, since it cleanly defines a 
homogeneous {\it mean field} alloy, it constitutes a natural starting 
point for perturbative studies of the fluctuations. The last feature 
has allowed the development of successful methods for the study of 
phase equilibria~\cite{phase} and magnetic phenomena~\cite{magnetic} in 
metallic alloys and led to surprisingly accurate calculations of 
properties connected with typical Fermi liquid effects, such as 
spin~\cite{spinwaves} and concentration waves~\cite{cwaves}.

Since 1990, in spite of all the above successes, the CPA theory, or, better, 
the way in which it has been first implemented self-consistently within 
the density functional theory~\cite{CPA-DF}, has been 
criticised because it does not account for the electrostatic energy that, in 
metallic alloys, arise from charge transfers~\cite{Magri}. Alternative 
models have been proposed able to cope with this lack~\cite{CWZ} and to 
determine more accurately total energies and mixing enthalpies. 
Unfortunately, the theories of this new class do not allow for an easy 
evaluation of spectral properties and are usually much more 
computationally demanding than the CPA.

In the last ten years several attempts have been made to improve the 
CPA theory including electrostatic interactions. The starting point of 
these modified CPA theories is the consideration of the physical mechanism 
responsible of the screening of impurity charges in metals, this is 
generally took into account introducing some {\it screening length} of 
the order of the nearest neighbours distance. Such theories, the screened 
impurity model by Abrikosov et al.~\cite{Abrikosov} or the screened CPA 
method by Johnson and Pinski~\cite{more_scr} or the more recent model 
by Zaharioudakis et al.~\cite{Zaharioudakis}, are able to achieve 
considerable improvements over the standard CPA, both for the total 
energies and the spectral properties~\cite{Abrikosov_cpa,isomorphous}.

In the last few years, the advent of order N electronic structure 
calculations~\cite{LSMS,LSGF}, making feasible the study of large 
supercells containing hundreds of atoms, has led to a very remarkable 
result concerning the charge transfers in metallic alloys. Namely, 
Faulkner, Wang and Stocks~\cite{FWS1,FWS2} discovered that local charge 
excesses in random alloys are related by simple linear relations, 
the '$qV$' laws, to the local electrostatic potentials. This new result, 
although not yet formally derived within the context of the density 
functional theory, has been confirmed by other order N 
calculations~\cite{Abrikosov_cpa} and, indirectly, by photoemission 
experiments~\cite{Weightman,FWS3} and does not appear compatible with 
previously proposed models~\cite{Magri,Wolverton}. It is interesting to 
observe that the existence of $qV$ linear relations has been obtained also 
within a Thomas-Fermi model~\cite{Pinski-TF} and, henceforth, does not 
appear related with the specific form of the density functional used, but, 
rather, to the achievement of self-consistency in density functional 
theory calculations along with the solution of the Poisson equation used 
for the potential reconstruction.

The new results of Refs.~\cite{FWS1,FWS2} changed the scenario: the 
local charge excesses take all the possible values over some range, 
they can be described by a distribution that cannot be reproduced just 
by considering the nearest neighbours environment. The standard CPA theory 
or its modified 
versions~\cite{Abrikosov,more_scr,Zaharioudakis}, cannot describe such a 
distribution, nor other kind of models~\cite{CWZ} can. A possible way out 
is to renounce to the assumption of a single atomic potential for each 
alloying species, here referred as to the {\it isomorphous assumption}. 
Thus, for instance, the {\it polymorphous} CPA (PCPA) of Ujfalussy et 
al.~\cite{Ujfalussy,notePCPA}, through the explicit inclusion of the 
electrostatic interactions between different crystal sites, leads to 
different site potentials. Though 
very successful from many points of view, however, the PCPA model, at 
similarity of order N calculations, is based on the use of supercells. 
As noticed by Faulkner~\cite{Faulkner}, this does not forbid to obtain 
accurate ensemble averages for the relevant physical observables, 
nevertheless the kind of averages that can be obtained from supercells 
is bound to some specified model of disorder, usually the random alloy 
model in which no correlation is assumed for the occupations of two 
different crystal sites. This, for instance, forbids the study of 
order-disorder phase transitions.

In this paper we shall follow an alternative route. We shall analyse 
the response of the {\it mean field} CPA alloy to external local fields 
designed to simulate the effects of local electrostatic fields. Our main 
result will be that the $qV$ relations can be obtained within 
the context of the isomorphous CPA theory, in satisfactory agreement with 
order N calculations. The theoretical scheme we shall develop 
allows for full charge relaxation at the impurity sites and clarifies 
that the linearity of the $qV$ relations holds well beyond the linearity 
regime for other site properties. Of course, in virtue of the 
fluctuation-dissipation theorem, the response to the external field that 
can be obtained from our theory must be equal to the response to the 
internal field due to the electrostatic interactions. Thus, our 
results could constitute a first step towards 
a CPA based theory that at the same time will be able to account accurately 
for the electrostatic interactions in metallic alloys and that maintain all 
the traditional strengths of the CPA. Another important feature of our work 
is that it does not rely on some specified supercell and, thus, in principle, 
can be used for studying ordering phenomena in metallic alloys.

The remainder of this paper is organised as follows. In the next Section 
II we shall analyse the information about charge transfers that can be 
obtained from the standard isomorphous CPA theory in the light of the 
new results from 
order N calculations. In Section III we shall describe a new version of the 
Coherent Potential Approximation that includes 
local fields (CPA+LF) as well as the applications of the model to fcc CuPd 
and fcc and bcc CuZn alloys. In 
the same Section III we shall discuss the picture obtained for the charge 
relaxation phenomena, connected with the screening of the external field. 
In the final Section IV we shall make our summary and draw our conclusions.

\section{Charge transfers in metallic alloys.}
\subsection{Charge transfers in the CPA theory.}
As noticed by Magri et al.~\cite{Magri}, the Coherent Potential 
Approximation~\cite{Soven} does not include the energetic contributions arising 
from the charge transfers between different sites in metallic 
alloys~\cite{note1}. This notwithstanding, sometimes, the information contained 
in the CPA theory about the same charge transfers, has been helpful for 
understanding the physical properties of certain systems~\cite{BGG,BG}. Below 
we shall elaborate about the reasons for this apparent paradox. 

\begin{figure}
\centerline{\epsfig{figure=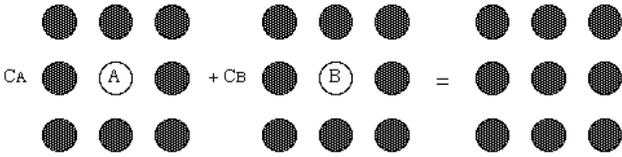,width=8.5cm}}
\caption{Schematic illustration of the CPA theory. Dark sites are 
occupied by the CPA coherent scatterer described by the single site 
scattering matrix $t_C$. The central impurity sites, labelled by A and B, 
are characterised by the single site matrices $t_A$ and $t_B$.}
\label{cpa}
\end{figure}

The CPA theory (we use the multiple scattering theory 
formalism~\cite{Gyorffy,KKRCPA}) deals with a random binary alloy $A_{c_A}B_{c_B}$ by 
solving for $t_C$ the so called CPA equation,
\begin{equation}
\label{CPA}
c_A G_A(t_A,t_C)+c_B G_B(t_B,t_C)=G_C(t_C)  
\end{equation}
The three Green's functions in Eq.~(\ref{CPA}), $G_A(t_A,t_C)$, 
$G_B(t_B,t_C)$ and $G_C(t_C)$, refer to the three different 
problems sketched in Fig.~\ref{cpa}. In fact, $G_C(t_C)$ is the Green's function 
for an infinite crystal whose sites are all occupied by effective scatterers
characterised by the single-site scattering matrix $t_C$. On the other 
hand, $G_{A(B)}(t_{A(B)},t_C)$ is the Green's function for a single impurity 
'atom' described by the single-site scattering matrix $t_{A(B)}$ and embedded in 
an infinite crystal with all the other sites  
characterised by the single-site scattering matrix $t_C$. 
While the homogeneous effective crystal, the 'coherent' medium of the CPA 
theory, let us call it C, is electroneutral, the two impurity Green's 
functions lead to net charge excesses, $q^0_A$ and $q^0_B$, in the sites occupied 
by the A and B impurities. On behalf of Eq.(\ref{CPA}), these charge excesses 
satisfy the condition,
\begin{equation}
\label{elecn}
c_A q^0_A + c_B q^0_B=0 
\end{equation}
However, in the coherent medium C, there is no charge transfer from one 
site to the others, thus, Eq.(\ref{elecn}) cannot be interpreted
as an ordinary electroneutrality condition. This notwithstanding, the excess 
charges, $q^0_{A(B)}$, can legitimately be considered as the amount of 
charge that the impurity site A(B) would attract from the medium 
C~\cite{note1}, as in the two configurations on the LHS of Fig.~\ref{cpa}.
We could think that Eq.~(\ref{elecn}) describes an {\it indirect} 
charge transfer, from A to C and from C to B. Since C is a mean field 
approximation that accurately accounts for the electronic properties 
of the alloy considered~\cite{Abrikosov_cpa}, we could say that the CPA 
'charge transfers' in Eq.(\ref{elecn}) {\it reflect} the {\it electronegativity 
differences} between A and B atoms in the random alloy $A_{c_A}B_{c_B}$
and, hence, contain valuable physical information. To put it into other 
words: in the physics of metallic alloys the {\it mean field metal} C  
plays the same role of {\it reference system} that, for molecules, 
is played by the H atom according to Pauling's celebrated concept of 
electronegativity~\cite{Pauling}. 
A further reason of interest for the CPA 'charge transfers', $q^0_{A(B)}$, is 
the fact that these, though obtained from an isomorphous model, at similarity 
of the 'true' 
charge transfers, show up dependencies on the alloy concentration that cannot be 
reproduced by trivial models~\cite{BGG}.

An additional difficulty in interpreting $q^0_{A(B)}$ as the 'true' 
charge transfers in a random alloy is that, on a more physical ground, the 
presence of a net charge on A should induce charges on the effective 
scatterers C and these, in turn, would affect the charge on A. Of course, 
as it is implied by its mean field nature, the CPA theory misses these 
important charge 
polarisation effects. It is easy to see that, the inclusion of these 
effects leads necessarily to a picture in which different sites occupied 
by atoms of the same kind are no longer equivalent because they are 
affected in different ways by the charge rearrangements. In such a case, 
the isomorphous 
approximation is no longer feasible. This circumstance led Ujfalussy et 
al.~\cite{Ujfalussy} to develop a {\it 'polymorphous'} coherent potential 
approximation (PCPA), in which different site potentials are allowed for. 

In this work we shall follow an approach alternative yet complementary to 
that of Ref.~\cite{Ujfalussy}. Borrowing the terminology by Ujfalussy et 
al., we shall include in the {\it isomorphous} CPA 
local perturbation fields and study the response of the charge density to 
these perturbations. 

\subsection{Charge transfers in 'exact' multiple scattering theory calculations.}
In the last few years, Faulkner, Wang and Stocks \cite{FWS1,FWS2} have extensively 
studied the distribution of charges that results from their Locally 
Self-consistent Multiple Scattering (LSMS)~\cite{LSMS} calculations in 
binary metallic alloys. LSMS calculations are basically exact, except for the 
Local Density and muffin-tin approximations used, and deal with large supercells 
containing 100 to 1000 atoms, designed to simulate substitutional disorder. The
principle result of Ref.~\cite{FWS1,FWS2} is that, for a given supercell 
configuration, a simple linear law relates the net
site charges, $q_i$, to the Madelung potentials, $V_i$, at the same sites, 
say,
\begin{equation}
\label{qvsv}
a_i q_i + V_i = k_i  
\end{equation}
For a binary alloy the quantities $a_i$ and $k_i$ take the values $a_A$ 
and $k_A$ if the i-th site is occupied by a A atom or $a_B$ and $k_B$ if 
it is occupied by B. 
The Madelung potentials $V_i$ depend on the charges at all the crystal sites, 
through the relationship
\begin{equation}
\label{Mad}
V_i = 2 \sum_i M_{ij} q_j 
\end{equation}
where the factor 2 comes from using atomic units. The Madelung matrix 
elements, $M_{ij}$, are defined~\cite{Ziman} as
\begin{equation}
\label{madmat}
\mbox{{\boldmath $M$}}_{ij}=\sum_{\mbox{{\boldmath $R$}}} \frac{1}{|
\mbox{{\boldmath $r$}}_{ij}+\mbox{{\boldmath $R$}}|}
\end{equation}
where the $\mbox{{\boldmath $r$}}_{ij}$ are the translations from the 
i-th to the j-th site within the supercell and $\mathbf{R}$ are the 
superlattice translation vectors. 

Very remarkably, no numerically significant deviations from the linearity 
have been found, at least in the range of the variations of $q_i$ and 
$V_i$ likely to occur in metallic alloys. This result, as stressed by 
the same authors~\cite{FWS1,FWS2}, is by no means obvious and should 
be understood as a consequence of the complex charge relaxation 
phenomena that occur, in the calculations, along with the achievement of 
self-consistency, or, in the physical world, through the  
mechanism mentioned in the previous subsection. Actually, it has been shown that 
simple models assuming the site charges proportional to the number of 
unlike neighbours~\cite{Magri} lead to site charges 
distributions~\cite{Wolverton} not compatible with Eq.~(\ref{qvsv}). 

Although the above mentioned $qV$ linear laws have been extracted from first 
principle calculations, their formal derivation within the density 
functional theory has not yet been obtained. Therefore, Eq.~(\ref{qvsv})
should be considered as an empirical law that holds, at least, 
within the basic approximation underlying  
the LSMS theory, i.e. the local density approximation for the density 
functional theory and the muffin-tin approximation for the crystal 
potential. On the experimental side, the validity of Eq.~(\ref{qvsv}) is 
supported by photoemission spectroscopy measurements on random 
alloys~\cite{Weightman,FWS3}. 

An aspect of the above phenomenology, crucial for our present 
concerns, is the fact that, for random alloys, the four constants in 
Eq.~(\ref{qvsv}) depend only on the site occupations and the supercell 
configuration~\cite{note2}. Moreover the same constants exhibit only minor 
variations for different configurations corresponding to the same mean alloy
concentration. Remarkably, the dependence of the above constants on the 
molar fractions does not appear easy to describe within simple models. 
All the above circumstances  
suggest us that the constants in Eq.~(\ref{qvsv}), that appear to depend 
mainly on the site chemical occupation, could probably be 
calculated within a isomorphous theory based on the CPA. Below we shall 
develop such an approach. 

\section{Response to local fields of the 'CPA alloy'.}
\subsection{The local field CPA (CPA+LF) model}
As discussed in the previous section, LSMS calculations suggest 
that the charge excess at the $i$-th site of a metallic alloy is 
determined, at least within the accuracy permitted by numerical 
errors, by the local Madelung field, $V_i$, and by the site occupation, 
say $A$ or $B$, for a given alloy concentration. Borrowing the language 
of the Ginzburg-Landau theory, we can think of the whole set of 
local charges, $q_i$, as the order parameter field. The  
phenomenology suggests a local view in which the local excess of charge, 
$q_i$, can be considered as the response to the local perturbation 
field, $V_i$. 
    
\begin{figure}
\centerline{\epsfig{figure=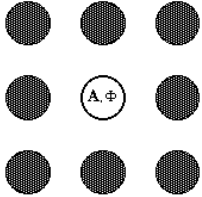,width=3cm}}

\caption{Schematic illustration of the CPA+LF method. As in Fig.~\ref{cpa}, 
dark sites are occupied by the CPA coherent scatterer described by $t_C$. 
In the central site, occupied by A, acts also a constant field $\Phi$.}
\label{cpaphi}
\end{figure}

In this section we develop a simple model for the charge response to such 
local perturbations. At variance of what happens in real systems, 
for which the local field at the i-th site is determined by the charges 
at all the other sites (see, e.g. Eq.~(\ref{Mad})), we shall treat the 
{\it external} local field, let us call it $\Phi$, as a parameter that 
can be varied at will. We imagine to have an A impurity atom in a otherwise 
homogeneous crystal with all the other sites occupied by C scatterers. We 
suppose that the single site scattering matrix of the CPA  medium, $t_C$, 
and its Fermi energy, $E_F$, have been determined by the CPA theory for the 
binary alloy $A_{c_A}B_{c_B}$. The local external field, $\Phi$, 
takes a constant value within the impurity site volume and is zero 
elsewhere. This situation is pictorially represented in 
Fig.~\ref{cpaphi}. To simplify our discussion we shall solve the problem using
the Atomic Sphere Approximation (ASA). However, the following 
considerations hold for any cellular method,
and, with minor modifications, also for the muffin-tin approximation.

We shall refer to the impurity A in the presence of the external field 
$\Phi$ as to $(A,\Phi)$. When $\Phi=0$, the site Green's function associated 
with it, $G^\Phi_A(t^\Phi_A,t_C)$, reduces to the usual CPA Green's function, 
$G_A(t_A,t_C)$. When $\Phi\ne 0$, $G^\Phi_A(t^\Phi_A,t_C)$ can be readily 
obtained using the multiple scattering theory impurity 
formula~\cite{KKRCPA}:

\begin{eqnarray}
\label{Green} \nonumber
G^\Phi_A(E, \vec{r}, \vec{r}^{\;\prime})  & = &
\sum_{L,L^\prime} [ Z^\Phi_L(E, \vec{r}) 
\tau^\Phi_{A,LL^\prime} Z^\Phi_{L^\prime} (E, \vec{r}^{\;\prime}) - \\
 & & Z^\Phi_L(E, 
\vec{r}) J^\Phi_{L^\prime} (E, \vec{r}^{\;\prime}) \delta_{LL^\prime}]           
\end{eqnarray}
where
\begin{equation}
\label{tau}
\tau^\Phi_A=D^\Phi_A \tau_C=\left[1+\tau_C \left( (t^\Phi_A)^{-1}-t_C^{-1} 
\right) \right]^{-1} \tau_C 
\end{equation}
In Eqs.~(\ref{Green}) and (\ref{tau}), E is the energy, $t_C$ and $\tau_C$ are 
the CPA single site scattering matrix and scattering-path operator, 
{\it as determined by an isomorphous CPA calculation}, i.e. $\Phi=0$, 
for the alloy at hand. The single site scattering matrix corresponding 
to $(A,\Phi)$, $t^\Phi_A$, is to be determined from the site 
potential $V^\Phi_A(\vec{r})+\Phi$, $D^\Phi_A$ is the CPA projector relative 
to the same site potential, $Z^\Phi_L(E, \vec{r})$ and $J^\Phi_L(E, \vec{r})$ 
are two orthogonal solutions of the Schroedinger equation for the same potential, 
chosen in such a way that the first behaves regularly at $r=0$. In our notation 
$L=(l,m)$ labels the angular momentum quantum numbers and, for sake of 
simplicity, we omit the energy dependence of all the scattering matrices. 
A complete account of the notation can be found in Refs.~\cite{KKRCPA}.

The charge density corresponding to $(A,\Phi)$ is obtained integrating 
Eq.~(\ref{Green}) over the energy to the Fermi level,
\begin{equation}
\label{rhoel}
\rho^\Phi_A(\vec{r})=-\frac{1}{\pi} Im \; \bigg\{ \int_{-\infty}^{E_F} dE \;
 G^\Phi_A(E, \vec{r}, \vec{r}^{\;\prime}=\vec{r}) \bigg\}            
\end{equation}
The corresponding site potential, $V^\Phi_A(\vec{r})$, can be reconstructed 
by solving the appropriate Poisson equation and adding the exchange-correlation 
contribution~\cite{Janak,Winter}. Unless $\Phi=0$, it will be 
different from the site potential obtained from the zero field CPA theory, 
$V_A(\vec{r})=V^{\Phi=0}_A(\vec{r})$, due to charge relaxations 
expected to screen in part the external field. 
In a numerical implementation of the theory, Eqs.~(\ref{Green}-\ref{rhoel}) and 
the potential reconstruction need to be iterated starting from a convenient 
initial guess, until convergence is achieved for 
$V^\Phi_A(\vec{r})$ or, equivalently, for $\rho^\Phi_A(\vec{r})$. Hereafter we shall 
refer to the above model as to the Local Field CPA (CPA+LF).

Once convergence is obtained for the charge density, the net charge on the 
site A 
can be obtained by integrating over the atomic sphere volume and subtracting 
the nuclear charge, $Z_A$,
\begin{equation}
\label{charge}
q_A(\Phi)=\int d\vec{r} \rho^\Phi_A(\vec{r}) - Z_A 
\end{equation}

It is important to realise that, while the above self-consistent 
determination of $V^\Phi_A(\vec{r})$ or $\rho^\Phi_A(\vec{r})$ allows for 
full charge relaxation at the impurity site, the CPA+LF does not modify the 
properties of the CPA medium C: these remain specified by the quantities $t_C$ 
and $E_F$ determined at zero external field. In other words, the charge relaxation 
that is likely to occur also in the neighbouring sites is not accounted for. 

Although in this paper we deal only with random alloys, we like to mention 
that our formalism can be applied also to the study of the screening phenomena 
in pure systems. In this case, of course, the CPA scattering matrices, 
$t_C$ and $\tau_C$, in Eq.~(\ref{tau}) should be replaced by their 
counterparts for a pure system.

\subsection{CPA+LF results for CuZn and CuPd alloys: the charge relaxation}
We have implemented the CPA+LF theory within our well tested KKR-CPA 
code~\cite{KKRAlgorithms}. If $t_C$ and $\tau_C$ from a previous standard KKR-CPA 
calculation are stored on a convenient energy mesh, the extra computational 
efforts required by the CPA+LF calculation are negligible.  

In this paper we discuss results for fcc CuPd and for bcc and fcc CuZn 
random alloys at several concentrations. In all the cases we have used the 
Local Density approximation (LDA) for the exchange-correlation 
potential~\cite{HK&KS}, the ASA approximation for the site potentials and 
the angular momentum expansions have been truncated at $l_{MAX}=3$. We have 
a fully relativistic treatment for core electrons and use a scalar 
relativistic approximation for valence electrons. For all the alloy systems 
considered in this paper, the lattice parameters have been kept fixed on 
varying the concentration. In particular, we set $a=5.5$ a.u. and $a=6.9$ a.u. 
for bcc and fcc CuZn, and $a=7.1$ a.u. for fcc CuPd. The reasons for this choice  
are, in first place, avoiding the consideration of the lattice parameter variations, 
not relevant for the purpose of this paper, and, in second place,  
allowing for an easier comparison with the results of Refs.~\cite{FWS1,FWS2}. For 
future reference, we mention that with this choice, the atomic volumes in 
fcc and bcc CuZn alloys differ only about 1.3 per cent.

\begin{figure}
\centerline{\epsfig{figure=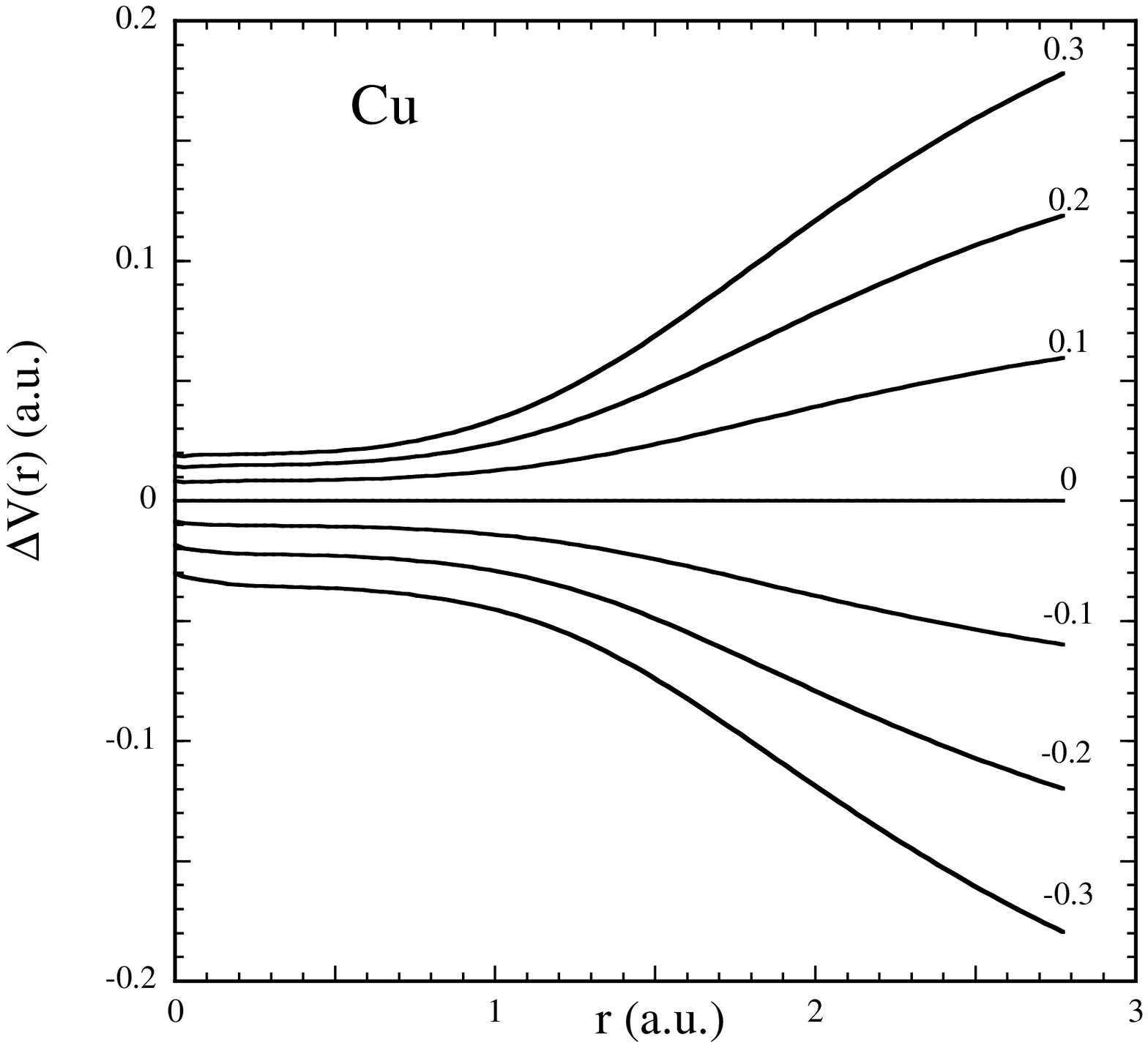,width=6.5cm}}
\centerline{\epsfig{figure=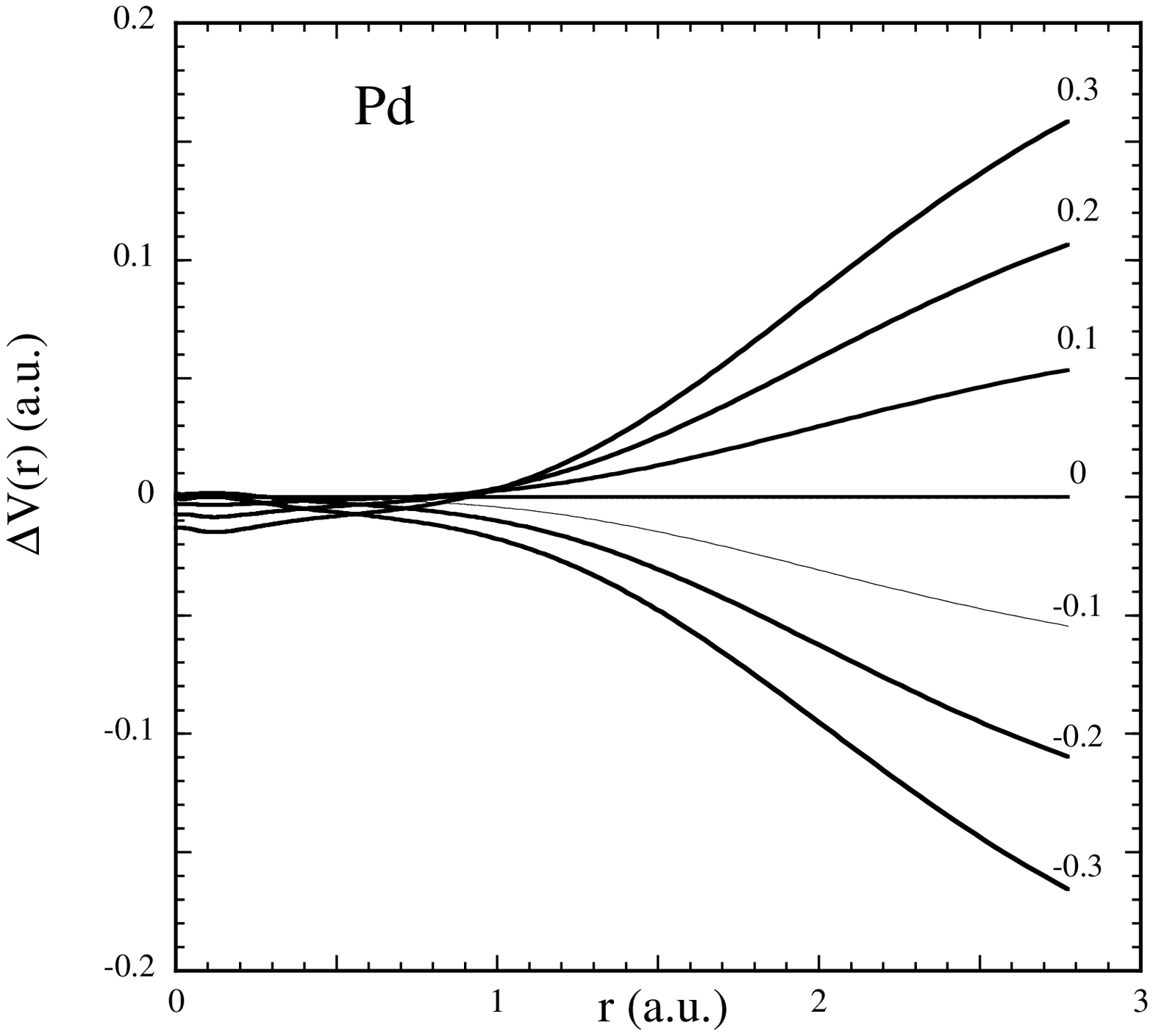,width=6.5cm}}
\caption{Calculated total field $\Delta V^\Phi_\alpha(r)$,$\alpha=$Cu, Pd 
(see Eq. 10) in Cu$_{0.50}$Pd$_{0.50}$ fcc random alloys.
The labels indicate the values of the external field, $\Phi$. At the 
Wigner-Seitz radius, $r_{WS} \approx 2.77$ a.u., the total field results to 
be about one half of the external field, while the electronic screening is 
almost complete at $r=0$.}
\label{vvsphi}
\end{figure}

As we said, the CPA+LF model allows for the determination of the response 
to an external potential field by the electrons {\it inside} the atomic sphere A. 
More specifically, the difference 
\begin{equation}
\Delta V^\Phi_A(\vec{r})=V^\Phi_A(\vec{r})+\Phi-V^{\Phi=0}_A(\vec{r})
\label{deltav}
\end{equation}
can be 
interpreted as the sum of the external field, $\Phi$, and the internal 
screening field inside the atomic sphere. Some typical trends for this 
quantity are shown in Fig.~\ref{vvsphi}. There we report $\Delta 
V^\Phi_\alpha(\vec{r})$, $(\alpha=Cu,Pd)$, for an fcc Cu$_{0.50}$Pd$_{0.50}$ 
random alloy, that we have selected as a typical case. At the Wigner-Seitz radius, 
$r_{WS} \approx 2.77$ a.u., the internal field is able to 
screen about one half of the external field, both for Cu and Pd impurities, 
while the screening is almost complete for $r<1$ a.u.. 
Apparently, the effect of the screening is far from being just a constant 
shift of the local chemical potential: if that was the case, in Fig.~\ref{vvsphi} 
we would have just equally spaced horizontal lines. What we observe 
is much more complicated. For instance, in the case of Pd 
impurities we see that, for the largest values of $\Phi$ considered, 
$\Delta V_{Pd}^\Phi(r)$ takes {\it negative} values at small r's, thus, in the 
case of Pd, the external field appears {\it overscreened} in the same range of 
$r$.

\begin{figure}
\centerline{\epsfig{figure=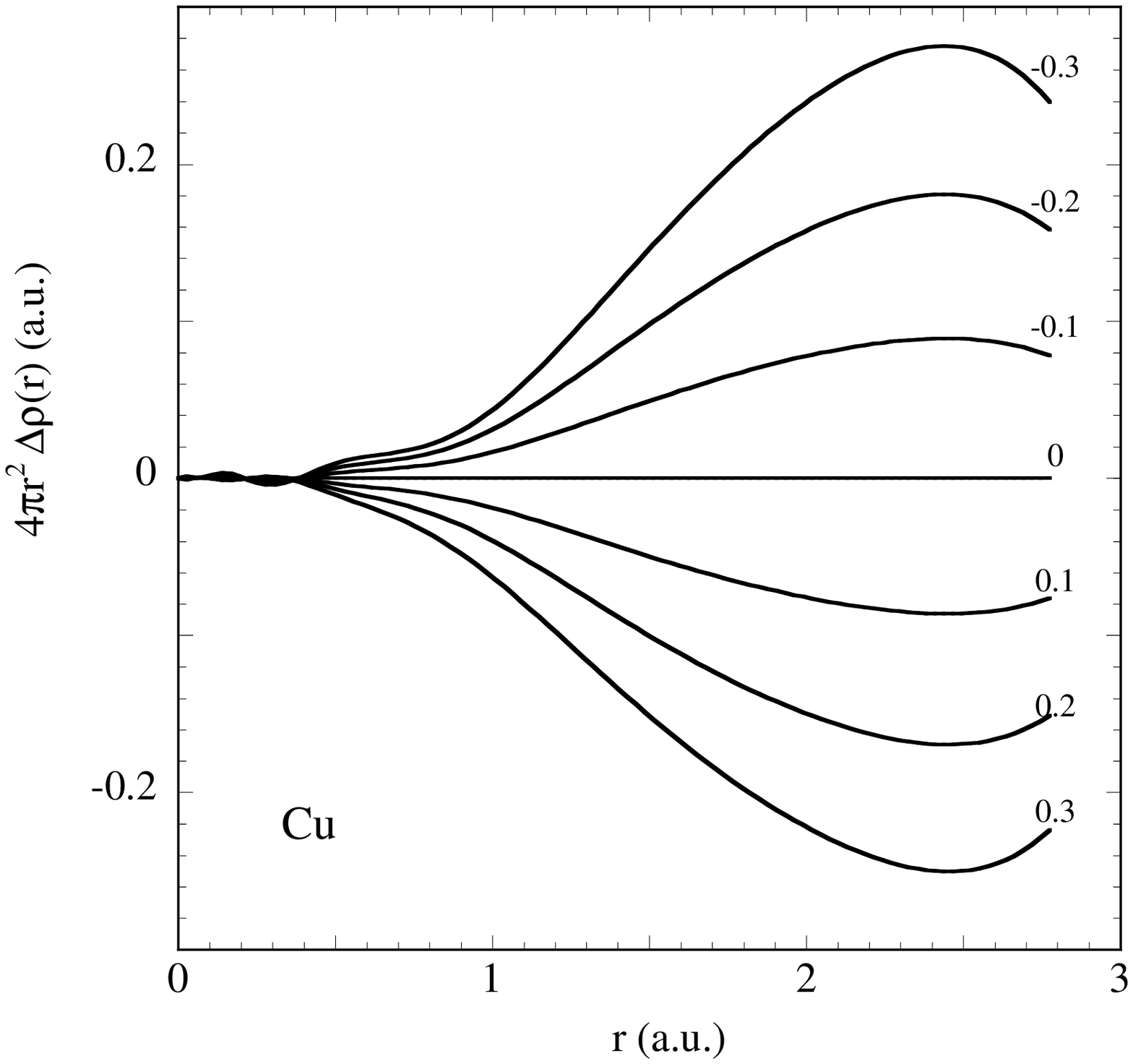,width=6.5cm}}
\centerline{\epsfig{figure=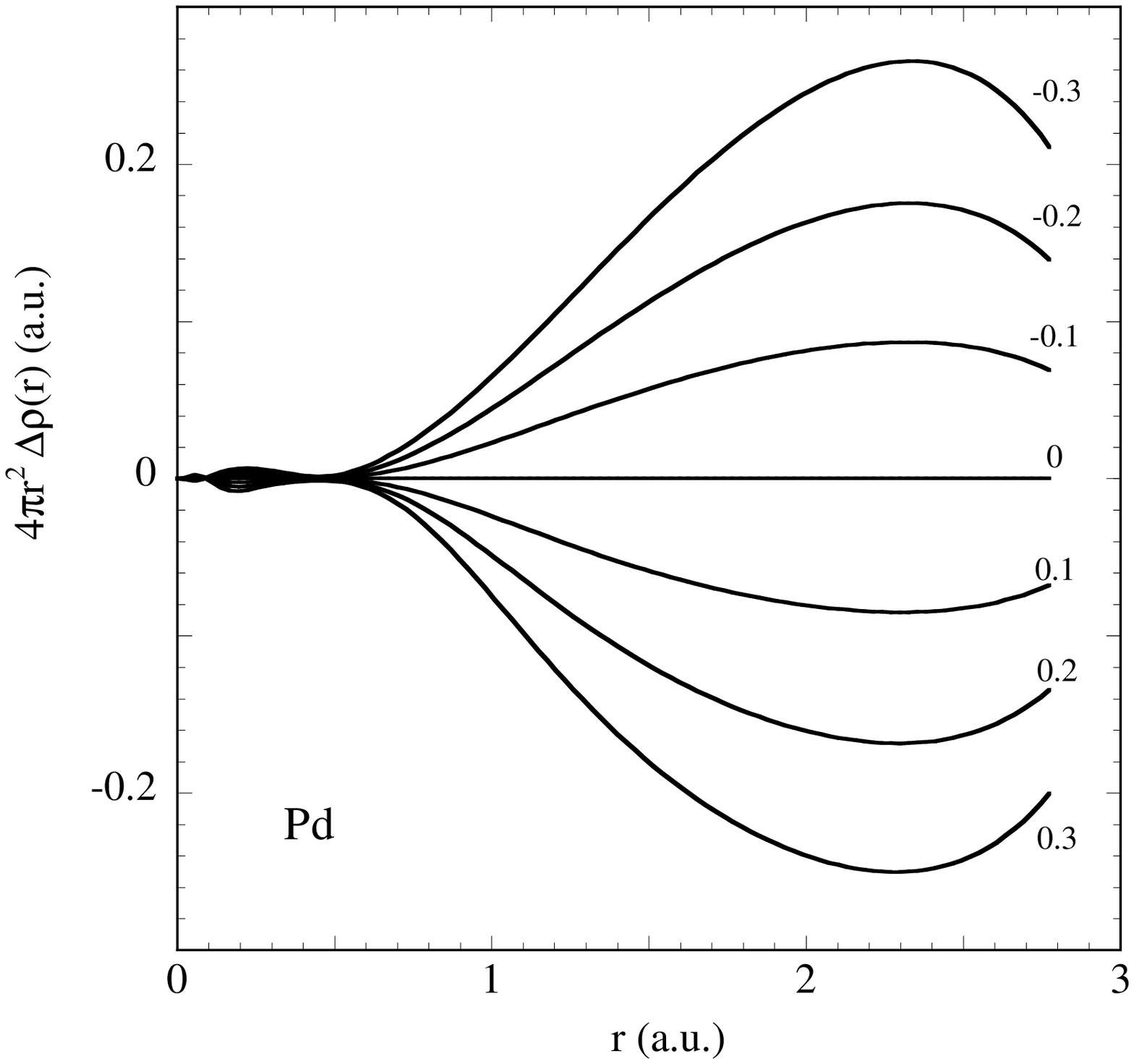,width=6.5cm}}
\caption{Calculated excess charge density, $4 \pi r^2 \Delta \rho^\Phi_
\alpha(r)$ ($\alpha=$Cu, Pd) (see Eq. 11) in Cu$_{0.50}$Pd$_{0.50}$ 
fcc random alloys. The labels indicate the values of the external field, $\Phi$.}
\label{rhovsphi}
\end{figure}

The complex nature of the screening phenomena is further confirmed by a look 
at the electronic densities. In Fig.~\ref{rhovsphi} we plot the excess 
charge density induced by the external field
\begin{equation}
\Delta \rho^\Phi_A(\vec{r})=\rho^\Phi_A(\vec{r})-\rho^{\Phi=0}_A(\vec{r})
\label{deltarho}
\end{equation}
 both for Cu and Pd sites, again for random fcc Cu$_{0.50}$Pd$_{0.50}$. The 
largest effects come from the large $r$ region, where the electron density 
decreases on increasing $\Phi$ (everywhere in this paper the 
expressions "electronic density" or "charge density" are used indifferently 
with the meaning of "electron number density", i.e. the charge factor, 
$-e$, is {\it not} included). In the innermost part of the atomic spheres, 
the variations of the charge density 
sometimes may have opposite sign with respect to that observed close to 
the cell boundary.

\begin{figure}
\centerline{\epsfig{figure=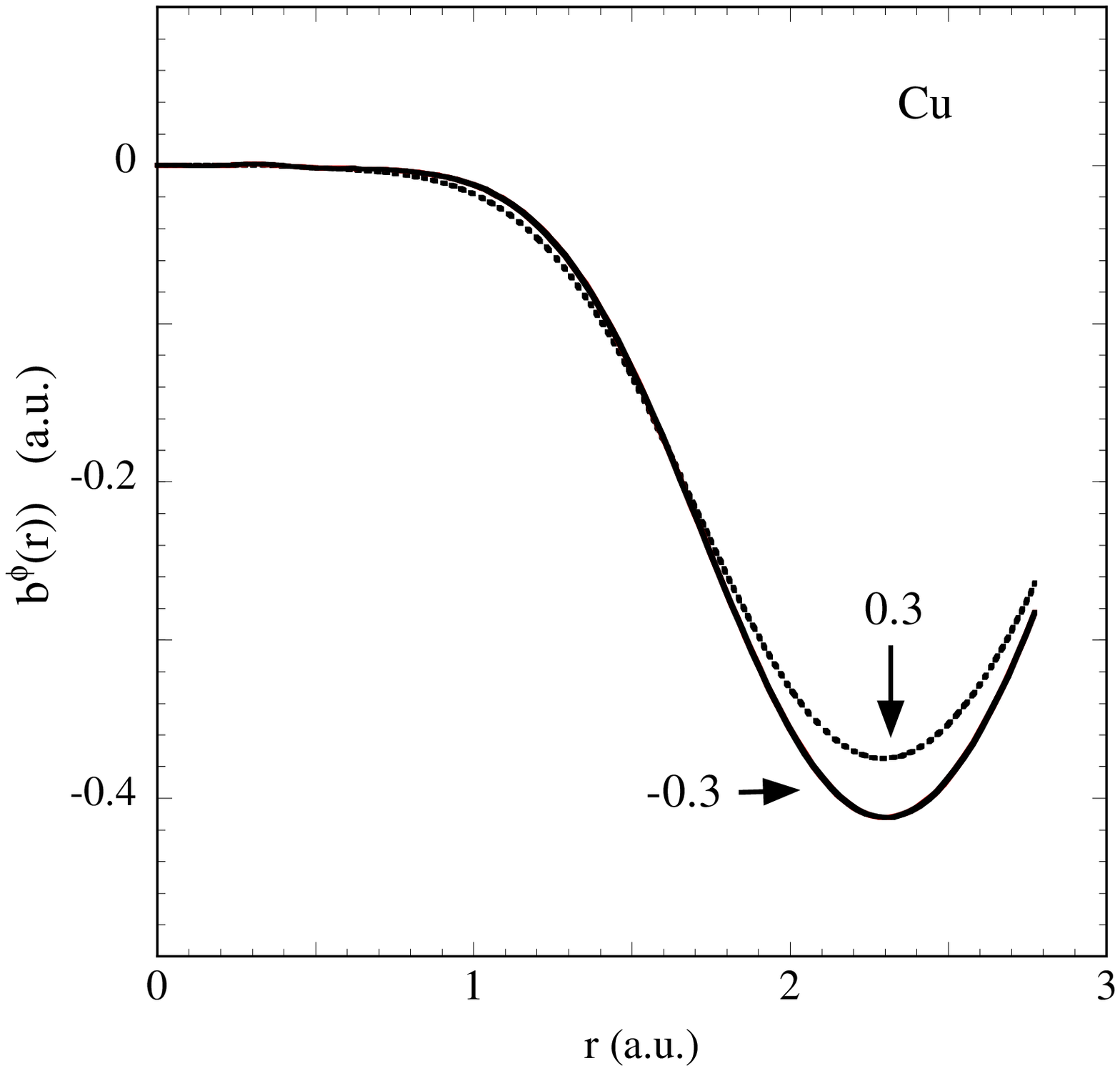,width=6.5cm}}
\centerline{\epsfig{figure=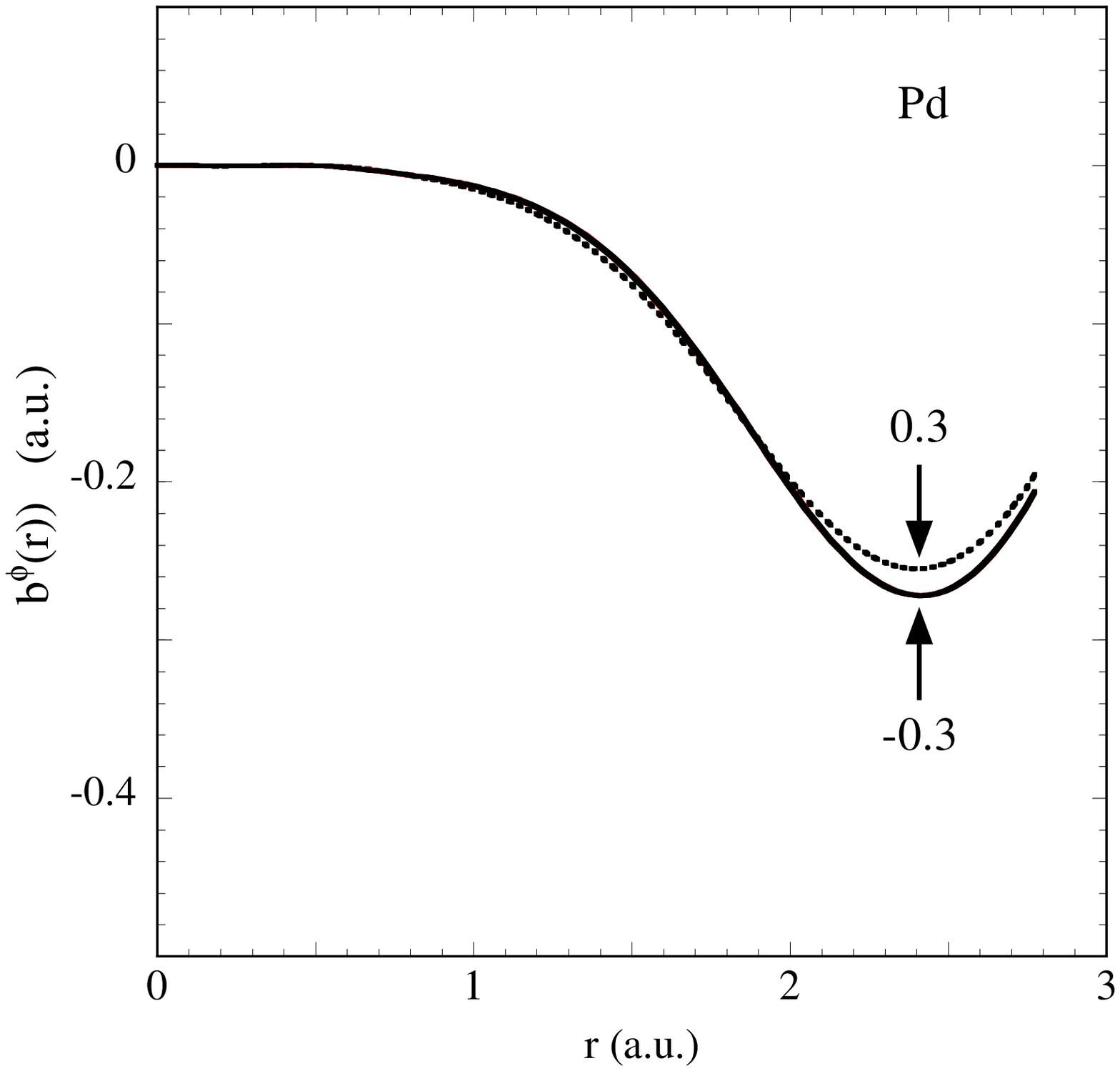,width=6.5cm}}
\caption{ The 'logarithmic derivative' with respect to the external 
field(Eq.(12)), 
$b^\Phi_\alpha(r)$ ($\alpha$=Cu,Pd) in Cu$_{0.50}$Pd$_{0.50}$ fcc random alloys. 
The continuous and the dotted lines refer, respectively, 
to $\Phi=-0.3$ and $\Phi=0.3$, i.e. the lowest and the highest $\Phi$ 
values considered in Fig.3.}
\label{rhologvsphi}
\end{figure}

We have considered also the quantity,  
\begin{equation}
\label{logder}
b_\alpha^\Phi(r)=\frac{\rho_\alpha^\Phi(r)-\rho_\alpha^{\Phi=0}(r)}
{\Phi \;\rho_\alpha^{\Phi=0}(r)} \approx \frac{\partial}{\partial \Phi} 
log \rho_\alpha^\Phi(r)
\end{equation}
that, in the limit $\Phi \rightarrow 0$ reduces to the logarithmic 
derivative of $\rho_\alpha^\Phi(r)$ and that, on the basis of a formal 
scattering theory analysis~\cite{unpub} is expected to have a weak dependence 
on $\Phi$. As we can see from Fig.~\ref{rhologvsphi}, the residual 
dependence on $\Phi$ is about a few per cent in a relatively small $r$ 
interval not far from $r_{WS}$ and less then 1 per cent in 
most of the atomic sphere. Although the information conveyed by 
Fig.~\ref{rhologvsphi} can be valuable for the purpose of improving the 
initial guesses for the charge densities, however the dependence of 
$b_\alpha^\Phi(r)$ on $r$ appears still quite complicated: 
$b_\alpha^\Phi(r)$ is very small at small $r$'s and takes its largest 
(negative) value at about 2.3 Bohr radii for Cu and 2.4 Bohr radii for Pd. 
In this region of the atomic sphere the charge density appears most
sensitive to the external field.

\subsection{CPA+LF results for CuZn and CuPd alloys: the site charges}
As discussed in the previous subsection, the charge relaxation phenomena 
occurring when the local field $\Phi$ is included in the isomorphous CPA 
theory are quite complicated. It is then surprising to see that the 
corresponding site charges, calculated from Eq.~\ref{charge}, exhibit much 
simpler trends.

\begin{figure}
\centerline{\epsfig{figure=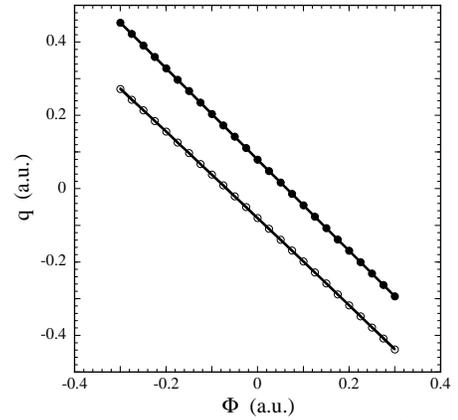,width=6cm}}
\caption{Site charge excesses $q_\alpha$ ($\alpha=$ Cu,Pd) vs. the external 
field, $\Phi$, from CPA+LF calculations for 
Cu$_{0.50}$Pd$_{0.50}$ fcc random alloys. Circles and diamonds, 
respectively, indicate Cu and Pd impurities.}
\label{qvsphi}
\end{figure}

In Fig.~\ref{qvsphi} we report $q_\alpha$ ($\alpha=$Cu, Pd) vs. $\Phi$ for 
a Cu$_{0.50}$Pd$_{0.50}$ fcc random alloy. As it is evident, the data can 
be fitted very well by two straight lines, one for each atomic species.
The resulting correlations differ from one by less than one part over a 
million. Interestingly, the slopes of the two lines are different by a 
relatively small but statistically relevant amount, about 5 per cent.
We notice that in Fig.~\ref{qvsphi} we have considered also $\Phi$ values  
considerably larger that those observed in LSMS calculation or 
likely to occur in real systems, thus our data support the view that the 
linearity observed has little to do with the size of $\Phi$. 

\begin{figure}
\centerline{\epsfig{figure=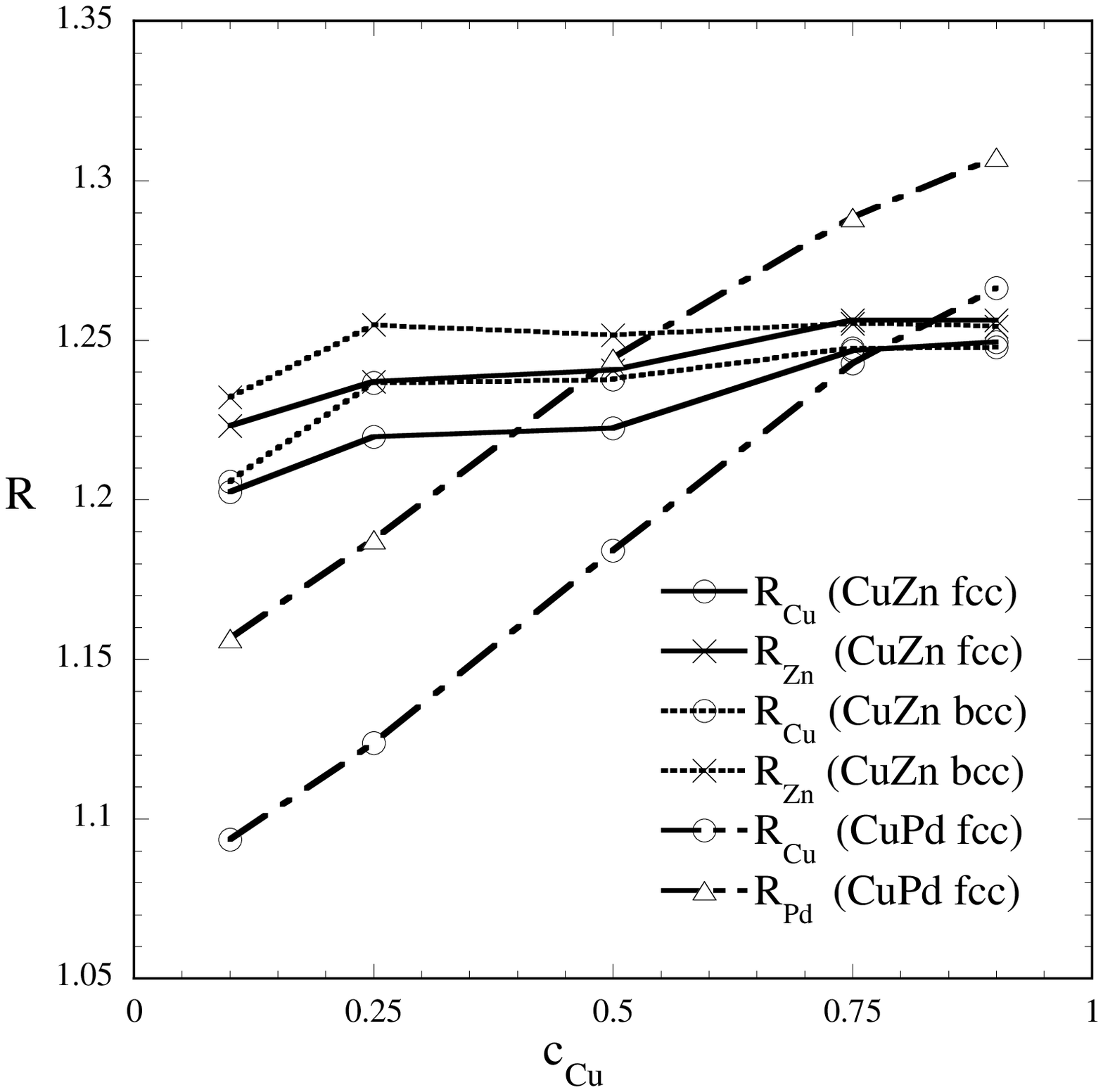,width=8cm}}
\centerline{\epsfig{figure=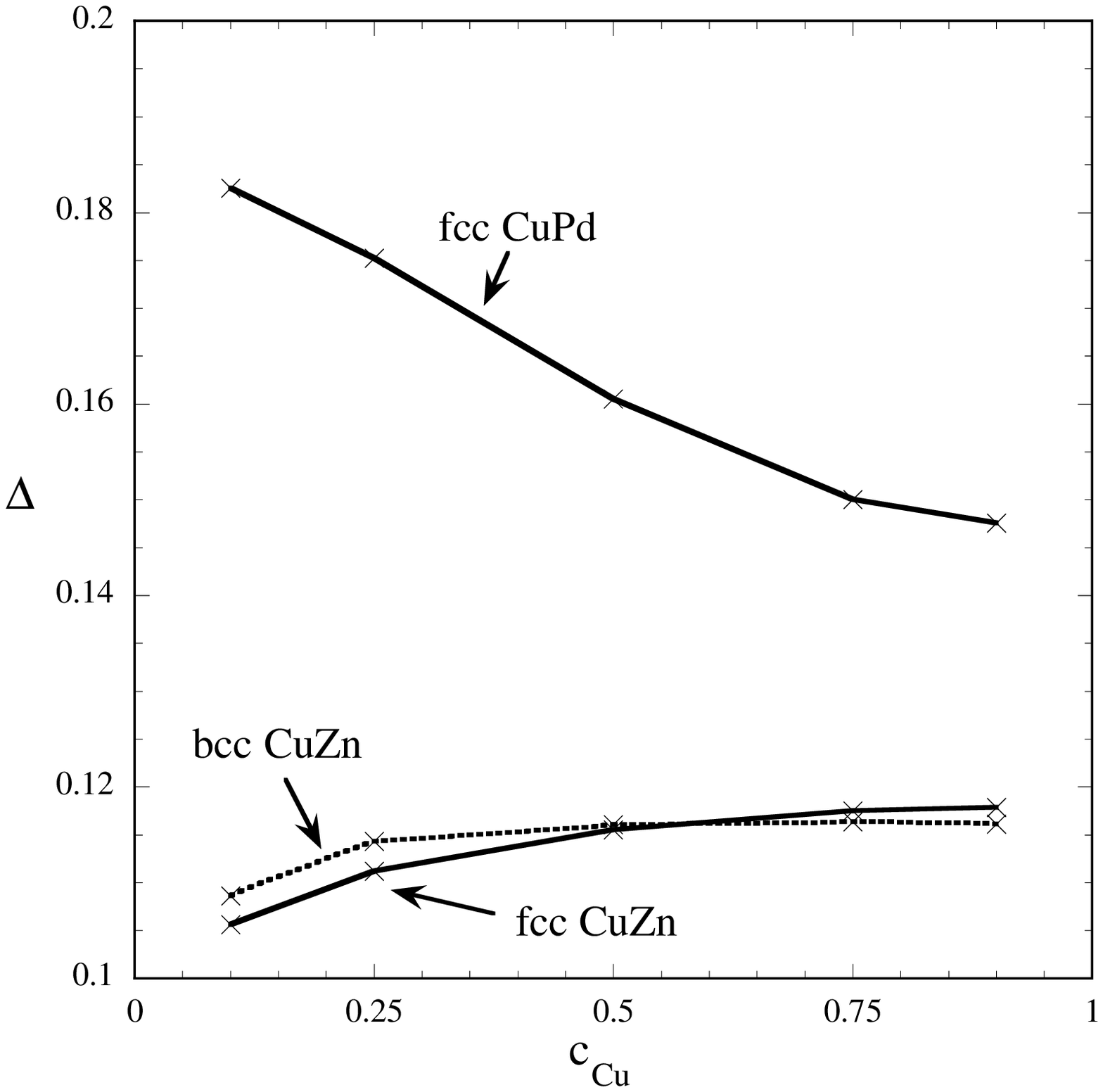,width=8cm}}
\caption{Fit coefficients of the linear law q vs.~$\Phi$ from CPA+LF 
calculations for fcc CuPd and fcc and bcc CuZn alloys plotted vs. the Cu 
content. Upper frame: response coefficients $R_\alpha$, ($\alpha$ refers 
to the alloying species); lower frame: 'electronegativity difference', 
$\Delta$. The various alloy systems are indicated by labels.}
\label{R_delta_vs_c}
\end{figure}

We have fitted the $q_\alpha$ vs. $\Phi$ curves at 
each molar fraction for fcc Cu$_c$Pd$_{1-c}$, fcc Cu$_c$Zn$_{1-c}$ and bcc 
Cu$_c$Pd$_{1-c}$ random alloys, at a number of alloy concentrations, using 
the linear relationships 
\begin{equation}
\label{qfit}
q_\alpha(\Phi)=q_\alpha^0 - R_\alpha \; \Phi
\end{equation}
However, at $\Phi=0$, our CPA+LF model satisfies the CPA 'electronegativity' 
condition, Eq.~(\ref{elecn}). Henceforth, $q_A^0$ and $q_B^0$ are not 
independent quantities and we have chosen as the parameters of our fit only 
the three quantities 
$R_A$, $R_B$ and 
\begin{equation}
\label{delta}
\Delta = q_A^0/c_B = - q_B^0/c_A = q_A^0  - q_B^0
\end{equation}
The results of these fits are reported in Table I. 
The deviations from linearity are always comparable with the numerical 
errors and the fits cannot be improved significantly using different 
functional forms. The trends found for the fitting parameters vs. the alloy 
molar fractions are shown in Fig.~\ref{R_delta_vs_c}. The dependence on 
the concentration is appreciable for all the fitting 
parameters, as expected on the basis of the arguments developed in Section 
II. Remarkably, the dependences on the alloy system and on the concentration 
appear at least as much important as that on the atomic species. Thus, for 
instance, for a given alloy system and concentration, there are relatively 
small differences between the values of $R$ corresponding to sites occupied 
by different atoms. On the other hand, we find much larger variations for $R_{Cu}$ 
throughout the alloy systems considered. It is 
interesting to observe that the trends for the slopes, $R_{Cu}$ and 
$R_{Zn}$, and for $\Delta$ are very similar in {\it both} fcc and bcc 
CuZn alloys. We notice also that $\Delta$, a measure of the electronegativity 
difference between the alloying species, exhibits, at least for CuPd 
alloys, non negligible variations vs. the concentration. In the model 
of Ref.~\cite{Magri}, the same quantity is assumed independent 
on the concentration. As we see from Table I, the values for $\Delta$ 
from our theory are systematically 25-35 per cent smaller than those from LSMS 
calculations. This is, actually, a feature of the standard CPA theory,  
because the local fields do not enter in the determination of 
$\Delta$ and it has already been discussed in the literature~\cite{BG}. This 
notwithstanding, the CPA is able to catch the qualitative 
trends of $\Delta$ vs. the concentration for all the systems considered.  

The LCPA+LF model gives for q vs. $\Phi$ the same 
linear functional form as that obtained for $q$ vs. $V$ from LSMS 
calculations. However, the differences between the two different sets 
of calculations forbid, at this stage, a direct comparison of the 
fit coefficients. In fact, as we have already stressed, our CPA+LF model 
does not account for charge relaxation outside the impurity site volume. 
By its construction, the CPA medium C is able to screen the impurity 
charge at $\Phi=0$, i.e. $q_\alpha^0$. We can think that this amount of 
charge is screened by the infinite volume of C. The introduction of the 
local field at the impurity site causes a local excess of charge, 
$q_\alpha(\Phi) - q_\alpha^0$,  with respect to the standard CPA. In order 
to have global electroneutrality in the CPA+LF theory, we have to 
introduce, somewhere outside the impurity site, an opposite amount 
of charge, $q_\alpha^0 - q_\alpha(\Phi)$. This can be accomplished
following the ideas contained in the screened impurity model (SIM-CPA) 
by Abrikosov et al.~\cite{Abrikosov} and 
supposing that the excess (with respect to the standard CPA) charge 
at the impurity site, $q_\alpha(\Phi)- q_\alpha^0$, is {\it completely 
screened} at some distance, $\rho$, of the order the nearest 
neighbours distance, $r_1$. Accordingly, each of the $n$ nearest 
neighbours of the impurity cell shall have, in the mean, a net 
charge excess $(q_\alpha^0-q_\alpha(\Phi))/n$. This, in turn, 
will cause an extra field $\Phi_1 = n (2/\rho)(q_\alpha^0-
q_\alpha(\Phi))/n=2(q_\alpha^0-q_\alpha(\Phi))/\rho$ on the 
impurity site. The total field at the impurity site will be then 
the sum of the external field $\Phi$ and of the above extra term, 
in formulae,
\begin{equation}
\label{screen}
V_\alpha =\Phi + 2 (q_\alpha^0-q_\alpha(\Phi))/\rho
\end{equation}
Then, by solving for $\Phi$ the last equation and substituting in 
Eq.~(\ref{qfit}), we find
\begin{equation}
\label{qfit_mod}
q_\alpha(\Phi)=q_\alpha^0 - 
\frac{R_\alpha}{1+2R_\alpha/\rho} \; V_\alpha = q_\alpha^0 - 
\tilde{R}_\alpha V_\alpha
\end{equation}

The coefficients $\tilde{R}_\alpha$ can be compared directly with 
the slopes of the $qV$ relations from LSMS calculations. 
However, the comparison, reported in Table I, requires a caveat: 
we have assumed $\rho=r_1$, i.e. a complete screening at the distance of 
the nearest neighbours. Actually, the screening lengths in metals are of the order 
of this distance~\cite{Pines}, but our estimate is too 
crude to expect for a very good quantitative agreement with LSMS calculations 
in which the charge relaxation is allowed at all the length scales. 
However, the agreement found is quite satisfactory, within 10 per cent, 
for CuPd alloys, while larger discrepancies are found for CuZn. 
Again, the trends for $\tilde{R}_\alpha$ vs. the 
concentration are qualitatively reproduced.

\section{Conclusions}
Using a convenient generalisation of the CPA theory to include a 
local external potential, we have been able to reproduce 
the peculiar linear relationship between the local charge excesses 
and the local Madelung fields, obtained previously from the 
analysis of LSMS data~\cite{FWS1,FWS2}. Our CPA+LF theory has a 
very simple structure and, as opposite to LSMS, requires modest 
computational efforts. The comparison of the linear laws coefficients 
with those from LSMS calculations is quantitatively satisfactory, 
though there is place for further improvement. In particular, it 
has been necessary to include some screening length, that we have 
kept fixed to the nearest neighbours distance, and that constitute 
the only non {\it ab initio} input of our theory~\cite{Abrikosov_cpa}. 
Work is in progress about this point, and we hope to present in the next 
future a fully {\it ab initio} version of the theory. 
We think that the CPA+LF model catches the basic physical mechanisms 
that determine the distribution of the charge excesses in metallic alloys and
we are currently working on a new version of the theory in which the 
distribution of local charge excesses and that of the local fields are 
obtained self-consistently. We are confident that this will allow for an 
accurate calculation of the electrostatic contribution to the total alloy 
energy and believe that these new techniques can prove useful for the 
study of the phase equilibria in metallic alloys. 

The main result of the present paper, having obtained the correct 
functional form for the $qV$ relations within a CPA-based theory, does not 
solve the problem of their derivation within the density functional theory. 
Nevertheless, our work offers a much more comfortable mathematical ground 
for the search of such a solution and clarifies two points that we like to 
list below.

i) A conceptual advantage of our CPA+LF theory is that it is build upon
a cleanly defined reference medium, the CPA alloy, that is kept fixed. 
Having fixed the reference medium, the source of the linear laws 
can be traced {\it only} in the CPA projectors and in the site 
wavefunctions (see Eqs.~(\ref{Green}) and (\ref{tau})), these, in turn, 
are determined by the nuclear charge on the impurity site and the 
coupling potential entering 
in the corresponding Schroedinger-Kohn-Sham equation: in other words, 
by the chemical species and the Madelung potential. This ensures that, 
for a given alloy, i.e. for some specified alloy Green's function, 
{\it any} site property depends {\it only} on the chemical species and 
the Madelung potentials. This applies also, to some extent, to the more 
realistic LSMS model in which, having fixed the alloy configuration, fixes 
the system's Green's function. Of course, in the LSMS model the Green's 
function is determined self-consistently, while in the CPA+LF model we use a 
reasonable approximation of it: this difference, however is not essential 
since in {\it both} cases there is a {\it unique} Green's function. A more 
important difference between CPA+LF and LSMS calculation is the fact that, 
in the first theory, the way the impurity site is embedded in the system is 
determined simply by the CPA projectors, while in a more complete 
treatment of the multiple scattering, as in the case of LSMS calculations, 
the relationships between the site-diagonal part of the Green's function at 
the impurity site and the full system's Green's function are more 
complicate. Also in the last case one could try to define something 
analogous to the CPA projectors, such generalised projectors, however will 
depend not only on the systems Green's function and on the impurity site 
properties but also on the environment of the neighbouring sites. The fact 
that the $qV$ laws have been obtained first from the analysis of LSMS data
suggests that corrections for the 
neighbouring sites are really important only for the electrostatic coupling 
term. This view is supported also by the very remarkable accuracy achieved 
by the polymorphous CPA theory~\cite{Ujfalussy}.

ii) Provided that the effect of the nearest neighbours is negligible, 
on the basis of the above point i), any site 
property is a unique function of the local field and chemical 
occupation. This, together with the linear response theory, of course, 
implies linear $qV$ relations for {\it small} fields. However, the 
range of fields investigated in this paper has been sufficiently wide 
to encounter non linear behaviours for many quantities, as, for 
instance, the value of the charge density at some specified $r$. 
Although the reasons for this remain elusive, we can conclude that 
the trends for $q$ remains linear {\it beyond the linear response 
regime} for $\rho(\vec{r})$. 

After the submission of the present work, two papers~\cite{Ruban1,Ruban2} 
appeared that treat some of the questions addressed here from a different 
point of view. In Ref.~\cite{Ruban1} it is demonstrated that linear  
$qV$ laws can be obtained within the SIM-CPA model of Abrikosov et 
al.~\cite{Abrikosov} {\it on varying the effective screening length} contained 
in the same model. Since the SIM-CPA assumes an effective field proportional 
to the charge excess, and the proportionality constant is the inverse of the 
above effective length, it is clear that the procedure described in 
Ref.~\cite{Ruban1} has effects similar to varying a local 
external field. Though we believe that the results of Refs.~\cite{Ruban1,Ruban2} 
offer an independent confirmation of ours, there are important differences 
between the two approaches. To mention but one: just because of its mathematical 
construction, the SIM-CPA model {\it forces} the $qV$ laws to have the {\it same} 
slopes for both the alloying species, while the LSMS calculations of 
Ref.~\cite{FWS1,FWS2}, in agreement with the CPA+LF calculations presented in 
this paper, find different slopes for different alloying species. 

\begin{acknowledgements}
We thank Professor J.S. Faulkner and Dr. Y. Wang for having made 
availiable in digital form the data of 
Refs.~\cite{FWS1,FWS2}. We acknowledge also discussions with Professor E.S. 
Giuliano.
\end{acknowledgements}

\twocolumn[\hsize\textwidth\columnwidth\hsize\csname @twocolumnfalse\endcsname
\begin{table}
\begin{tabular}{cccccccc|ccc}
Alloys & $c$  & $\Delta$ & R$_{Cu}$ & R$_{X}$ & RMS $\times 10^4$ & 
$\tilde{R}_{Cu}$ & $\tilde{R}_{X}$ & 
$\Delta$ & $\tilde{R}_{Cu}$ &  $\tilde{R}_{X}$  \\
\hline
fcc Cu$_c$Pd$_{1-c}$ & 0.10 & 0.183 & 1.093 & 1.156 & 1.8 & 0.762 & 0.792 & 0.238 & 0.833 & 0.843    \\
& 0.25 & 0.175 & 1.124 & 1.187 & 2.1 & 0.776 & 0.806 & 0.229 & 0.838 & 0.851    \\
& 0.50 & 0.160 & 1.184 & 1.244 & 1.9 & 0.805 & 0.832 & 0.219 & 0.843 & 0.851    \\
& 0.75 & 0.150 & 1.243 & 1.288 & 2.4 & 0.831 & 0.851 & 0.212 & 0.838 & 0.853    \\
& 0.90 & 0.148 & 1.267 & 1.307 & 4.4 & 0.842 & 0.860 & 0.211 & 0.836 & 0.853    \\
\hline
bcc Cu$_c$Zn$_{1-c}$ & 0.10 & 0.109 & 1.206 & 1.232 & 10  & 0.800 & 0.812 & 0.155 & 0.536 & 0.581  \\
& 0.25 & 0.114 & 1.237 & 1.255 & 10  & 0.814 & 0.822 & 0.159 & 0.526 & 0.554  \\
& 0.50 & 0.116 & 1.237 & 1.251 & 6.9 & 0.814 & 0.820 & 0.156 & 0.545 & 0.549  \\
& 0.75 & 0.116 & 1.247 & 1.255 & 5.0 & 0.819 & 0.822 & 0.155 & 0.567 & 0.564  \\
& 0.90 & 0.116 & 1.248 & 1.254 & 3.2 & 0.819 & 0.822 & 0.158 & 0.582 & 0.577  \\
\hline
fcc Cu$_c$Zn$_{1-c}$ & 0.10 & 0.106 & 1.202 & 1.223 & 8.2 & 0.805 & 0.815 & 0.145 & 0.575 & 0.628  \\
& 0.25 & 0.111 & 1.220 & 1.237 & 8.1 & 0.813 & 0.821 & 0.150 & 0.580 & 0.618  \\
& 0.50 & 0.116 & 1.222 & 1.241 & 5.5 & 0.814 & 0.822 & 0.151 & 0.600 & 0.622  \\
& 0.75 & 0.117 & 1.247 & 1.256 & 5.2 & 0.825 & 0.829 & 0.150 & 0.615 & 0.632  \\
& 0.90 & 0.118 & 1.249 & 1.256 & 3.3 & 0.826 & 0.829 & 0.152 & 0.616 & 0.630  \\
\end{tabular}
\label{tabI}
\caption{Fit parameters for the q vs. $\Phi$ relationships from CPA+LF 
calculations in fcc Cu$_c$Pd$_{1-c}$, bcc Cu$_c$Zn$_{1-c}$ and fcc Cu$_c$Zn$_{1-c}$ 
random alloys. The 'electronegativity 
difference', $\Delta$, and the response coefficients, $R_{Cu}$ and $R_X$, 
($X$ = Pd or Zn, as convenient) are defined in Eqs.~(13) and (15), 
RMS is the root mean square deviation. The 'renormalised' response coefficients, 
$\tilde{R}_{Cu}$ and $\tilde{R}_{X}$, are defined in Eq.~(17). On the right 
"we report $\Delta$, $\tilde{R}_{Cu}$ and $\tilde{R}_{X}$ from the LSMS 
calculations of Refs.~[19-20].}
\end{table}
]

\newpage 

\newpage

\begin{references}
\bibitem{note1}
The meaning of the world 'site' throughout this paper need some 
clarification. We refer to cellular schemes in which the volumes 
corresponding to each site sum up to the crystal volume. A different 
situation occurs in the case of the muffin-tin approximation, because 
the muffin-tin spheres do not sum up to the crystal volume. 
Thus, there is no contradiction with the fact that in CPA (as in pure systems)
muffin-tin calculations the site-potentials and the total energy contain terms 
due to the charge transfer from the muffin-tin volumes to the 
interstitials.
\bibitem{note2}
With the expression "supercell configuration" we always mean the set of the 
occupations numbers for each site of the supercell.
\end{references}
\end{document}